# Ultra-soft Thermal Diodes Enabled by Dual-Alkane-Based Phase Change Composites

Yunsong Pang,[1],*,# Junhong Li,[1],* Zhibin Wen,[1] Ting Liang,[2] Shan Gao,[1] Dezhao Huang,[3] Rong Sun,[1] Jianbin Xu,[2],# Tengfei Luo[4,5],# and Xiaoliang Zeng[1],#

[1]Shenzhen Institute of Advanced Electronic Materials, Shenzhen Institute of Advanced Technology, Chinese Academy of Sciences, Shenzhen 518055, China

[2]Department of Electronics Engineering, The Chinese University of Hong Kong, Shatin, N.T., Hong Kong 999077, China

[3]School of Power and Mechanical Engineering, Wuhan University, Wuhan, Hubei, 430072, China

[4]Department of Aerospace and Mechanical Engineering, University of Notre Dame, Notre Dame, Indiana 46556, United States

[5]Center for Sustainable Energy at Notre Dame, Notre Dame, Indiana 46556, United States

*These authors contributed to the work equally and should be regarded as co-first authors.

#Corresponding

Yunsong Pang, E-mail: ys.pang@siat.ac.cn,

Jianbin Xu, E-mail: jbxu@ee.cuhk.edu.hk

Xiaoliang Zeng, E-mail: xl.zeng@siat.ac.cn

Tengfei Luo, E-mail: tluo@nd.edu

## Abstract

Thermal diode, a type of device that allows heat to flow in one direction preferentially, can be employed in many thermal applications. However, if the mechanical compliance of the thermal diode is poor, which prevents its intimate contact with heat source/sink surfaces, the thermal rectification performance cannot be used to its full extent. In this work, we introduce a heterojunction thermal diode made of a phase change material (PCM) consisting of dual alkanes (hexadecane and paraffine wax) and polyurethane. The fabricated thermal diode exhibits an ultra-soft mechanical feature, with a low elastic modulus of 0.4 KPa and > 300% elongation until failure – the best values reported to date for thermal diodes. The measured thermal rectification factor is as high as 1.42 – in line with the theoretical model prediction. Molecular dynamic simulations reveal that the thermal rectification mechanism of the PCM-based thermal diode originates from the crystal-amorphous phase transition of the hexadecane terminal as the temperature bias flips. Therefore, the heat flow in the forward direction is greater than the flux in the reverse direction. A series of experiments and finite element analyses are employed to verify the feasibility of thermal diodes for applications. Our results demonstrate that the fabricated thermal diode can be potentially used in building envelop to help with temperature regulation and thus reduce energy consumption for space cooling/heating.

## Introduction

Decarbonization through energy saving is one of the most important routes to fight climate change. 23%-32% of the global energy consumption is attributed to building cooling and heating.[1, 2] Building envelops with adaptive thermal conductance are going to critical in reducing the building energy consumption.[3, 4] A study showed that a variable conductance building envelop could lead to energy savings anywhere from 7% to 42%.[5] In order to achieve temperature self-regulation, it is preferred that the material used for the building envelop have the ability to preferentially allow heat to transfer in one direction more than in the inverse direction.[3] Thermal diodes are such materials.[6-8] The performance of a thermal diode can be evaluated by the rectification factor, $\varepsilon$, which is quantified as the ratio of heat flux in the forward to that in the reverse direction:[9, 10]

$$\varepsilon = \frac{J^+}{J^-} \qquad (1)$$

By surveying the state of the art of macroscopic thermal diodes, which are most practical for thermal engineering (**Figure 1A**),[9, 11-34] we find that pure or composite phase change material (PCM)-based heterojunctions tend to have high thermal rectification factors. Such heterojunction thermal diodes are usually made of a stack of two dissimilar materials with at least one of them made of PCM. During operation, due to the PCM's temperature-dependent thermal conductivity, the effective thermal conductivity changes depending on the phase of the PCM which is influenced by whether it is exposed to high or low temperature. Alkane as a PCM has been explored for thermal diode applications[35] because of its phase transition near room-temperature [36, 37] and the relatively large change in thermal conductivity between the solid and liquid phases. Although a few studies have shown that PCM-based thermal diodes are feasible,[17-19] the PCM behavior cannot be well-controlled when it liquifies. The leakage of the PCM in the liquid state or the voids emerging at the interface due to the phase transition-induced volume shrinkage would render the thermal diode ineffective.[11, 38] Another disadvantage of the current state of the art thermal diode is that they have a high moduli in the range of $10^3 \sim 10^{12}$ Pa. Although some of them have relatively high thermal rectification factors, when mounted between the heat source and sink acting as thermal interface materials, the high modulus will lead to high contact thermal resistance,[39, 40] negatively affecting the heat flux passing through. Therefore, a thermal diode with a soft mechanical character is desirable.

Here, we propose a strategy to employ composites of polyurethane (PU) and alkane materials (hexadecane (C16) and paraffine wax) to form a thermal diode heterojunction. PU was chosen because its versatility in a wide range of coatings, adhesives, thermoplastic elastomers and composites applications.[41] PU can serve as a matrix to hold the PCM to overcomes the leakage problem observed in the literature.[42] Through modelling to optimize the thermal diode design, the rectification factor of our fabricated thermal diode reaches ~1.42, enabled by an alkane loading ratio of 80 wt.%. The fabricated thermal diode exhibits an ultrasoft feature compared to other similar types of thermal diodes (**Figure 1A**), which can resist 300% tensile strain and has an elastic modulus of 0.4 KPa, by far the best values reported to date for thermal diodes. Therefore, it can form effective contacts with rough surfaces. Molecular dynamic simulations are utilized to understand the rectification mechanism of our PCM-based thermal diode. It is found that when the C16 terminal undergoes phase transition, the heterojunction transitions from the crystal (wax)-crystal (C16) to crystal (wax)-amorphous (C16) state. This transition leads to the changes in the effective thermal conductance hence the thermal rectification effect. Finally, we modeled the benefit of our thermal diode as a building envelope, which is shown to resist heat flow from outside during hot daytime and release heat outward during cool nighttime.

**Results and Discussion**

As mentioned above, we combine the inclusion of alkane materials with the PU matrix to fabricate a composite thermal diode. Briefly, two types of alkanes, C16 and paraffine wax, are respectively well mixed with the PU precursor, which is then thermally cured at a high temperature (80 °C) to form the C16/PU and wax/PU composite gel terminals. The formed terminals are then laminated together to obtain the thermal diode heterojunction (**Figure 1B**). To prevent material failure, additional curation at a high temperature is performed to enhance the adhesion strength. The C16/PU terminal exhibits an apparent phase transition when the ambient temperature is changed around the melting point of 18 °C for C16 (**Figure 1C**). Then, samples are characterized using the differential scanning calorimetry (DSC), atomic force microscopy (AFM), uniaxial tension/compression, long focal microscope, and wide-angle X-ray diffraction (WXRD).

The rectification performance of the fabricated thermal diode is measured by an in-house testing platform (**Figure 2A**). The heat fluxes along the forward and reverse directions of a representative composite thermal diode with an alkane loading ratio of 80 wt.% are shown in **Figure 2B**. The heat fluxes increase with an increment of the temperature bias ($\Delta T = 10, 15, 20 \cdots 40$) established by the temperature set of $T^* \pm \frac{1}{2}\Delta T$ for heater and cooler, where $T^*$ is the phase transition temperature of the phase change terminal material (here is 18.8 °C of the C16/PU composite). Details are provided in the **Experimental Section**. The rectification factors of the thermal diode with various alkane loading ratios are plotted in **Figure 2C**. As the alkane loading increases, the thermal rectification factor becomes greater, which is as expected since the origin of the rectification is from the phase change of the PCM. The rectification factor can be up to 1.2 with 80 wt.% alkane loading ratio and 40 K of temperature bias.

However, 1.2 is relatively low to other thermal diodes reported in the literature. To find the reason, a one-dimensional heat conduction model is used to analyze the temperature profile along the heat transfer direction (**Figure S1** in **Supporting Information (SI)**). For the case of 40 K temperature bias along the reverse direction, the temperature at the contacting interface is below the phase change point (18.8 °C) of the C16/PU terminal as determined using DSC (**Figure S2A**). Hence, the C16 terminal is partially solid and partially liquid, and its effective thermal conductance is higher than that of the pure liquid phase. A similar phenomenon can happen to the forward case, resulting in lower effective thermal conductance and lower ultimate thermal rectification.

Since the phase change point of the C16/PU composite terminal with various loading ratios is a constant (~18.8 °C) and the enthalpy is linearly correlated to the alkane loading, the PU matrix is not expected to affect the intrinsic thermal properties of alkane (**Figure S2B**). Therefore, we can estimate the effective thermal conductivity ($k_e$) of the composite terminals with various loading ratios using effective medium approximation: [9]

$$k_e = \frac{k_{in} + 2k_m + 2\Phi(k_{in} + k_m)}{k_{in} + 2k_m - \Phi(k_{in} + k_m)} \tag{2}$$

where $k_{in}$ is the thermal conductivity of the alkane inclusion, and $k_m$ is the thermal conductivity of the PU matrix (see **Table S1**), and $\Phi$ is the alkane loading ratio. According to the above equation, **Figure 3A** displays the $k_e$ of C16 composite terminal when the $k_m$ is in the range of 0 to 0.4, and the loading ratio between 0 to 100%. Although our $k_m = 0.11$ W/mK for PU is fixed, we scanned $k_m$ in a range of possible polymer thermal conductivity to provide information for future studies where matrix materials other than PU may be employed. The effective thermal conductivity of the wax composite can be obtained by the same method.

Combing with the estimated $k_e$ of composite terminals, we can predict the highest rectification factor of the fabricated thermal diode by applying the model developed by Cottrill et al.[9, 43] since such model can completely describe the relationship between the thermal diode rectification factor, phase transition point, terminals' thickness ratio, and the set temperatures for cold and hot sides (details are provided in **SI**). **Figure 3B** shows the relationship between the rectification factor, thickness ratio, and reduced temperature, which are the three essential terms in the theoretical models. For the reduced temperature, $\bar{T}$ is defined as $\bar{T} = \frac{T_2 - T^*}{T^* - T_1}$, where $T_1$ and $T_2$ are the temperatures set on the code and hot surfaces respectively. **Figure 3C** shows that various temperature pairs can lead to different rectification factors based on the optimal thickness ratio of 6/4. Thus, according to this contour plot, we can optimize the rectification factor by adjusting the temperature pair ($T_1$, $T_2$). Besides this, the highest thermal rectification factor can also be estimated roughly with the given set of effective thermal conductivities above and below the C16/PU terminal's transition point, $k_1$ and $k_2$, respectively, as shown in the equation of $\varepsilon = \sqrt{k_1/k_2}$.[9] Based on this equation, we plot the effects of the loading ratio and $k_m$ on the rectification factor in **Figure 3D**. Using the thermal properties data of C16, PU, and loading ratios, the rectification factor is predicted to be 1.41, which closes to the calculated results from **Figure 3B** and **C**.

Based on the modeling results, we use the thickness ratio of C16 composite/wax composite = 6/4 (**Figure 3B**) and set the cold and hot side temperatures according to **Figure 3C**, and we measure the rectification factors of the thermal diode again. All the measured factors are within 10% of the predicted maximum of 1.42 and the peak value can be achieved as 1.405, when is under the temperature pair of 4 and 27 °C (**Figure 2E**). We also test the rectification factor of a thermal diode with a 70% alkane loading ratios, and the measured peak value is within 5% of the model prediction (**Figure S4**).

Additionally, we have tested the cycling stability of the thermal diode. As shown in **Figure 3F**, the forward and reverse heat flux shows no systematic drifting during the tested 40 cycles, and the thermal rectification is stable at ~1.38 with a standard deviation of ~0.01. Such a good cycling stability can be attributed to confinement of the alkane provided by the PU aerogel matrix. Conversely, if a pure PCM that without matrix is utilized, the fluidity of the material after the phase transition will lead it to leak during usage, causing the material to fail, as has been demonstrated by others.[44, 45] To show that the alkanes are well loaded to the PU matrix, we used the probe of AFM to scan the material modulus along the path from the matrix phase to inclusion as the schematic shows in **Figure 4A**. From the testing curves, there are two plateaus at the beginning and end of the path that refer to the modulus of matrix (PU) and inclusion (alkanes) respectively, and the middle parts represent the mixture phase modulus. The gradient change in modulus implies the absence of a distinct interface between the PU and alkanes. They are blended forming transition layers with certain thicknesses (~300 nm for PU and wax, ~100 nm for PU and C16). Such layers indicated that there is a good interaction between the PU and alkanes, thereby ensuring the confinement to make the material stable during the usage. The good interaction of PU and alkanes is due to the hydrophobic nature of the PU. Based on the contact angle measurement and subsequent calculation, the PU surface energy is 13.78 mJ/m$^2$, indicating hydrophobicity that enables it to mix well with alkane without phase separation (details are provided in **Experimental Section**).

The fabricated thermal diode is also mechanically stable as it does not fail after 1000 cycles of tensile tests (**Figure 4B).** It also exhibits an ultrasoft mechanical character, and the modulus decreases with the increasing alkane loading ratio for both tension (**Figure 4C**) and compression (**Figure 4D**) responses. The modulus reaches ~0.01 MPa when the alkane of loading ratio increases to 80 wt.% (**Figure 4E**). By increasing the alkane loading from 50 wt.% to 80 wt.%, the softness of the alkane gradually dominates the mechanical behavior of the entire thermal diode. For the tension test, the stress-strain curve is plotted with two segments: from the origin to the fracture point of the C16/PU terminal and from the fracture point of C16/PU to the wax/PU terminal one. The distinct mechanical behavior of the thermal diode confirms that the wax/PU terminal exhibits higher mechanical strength and more prolonged fracture tensile strains, which indicates that the wax/PU is mainly responsible for the mechanical performance of the thermal diode. It can be explained that only the PU matrix contributes mechanical strength when it composites with the liquid phase of C16. Moreover, increasing the loading ratio can defeat the mechanical strength of material since PU will lose domination to contribute to the supporting structure. We also employ the shore 00 rate as an indicator to evaluate the material's softness (**Figure 4F**), and the results show the same trend as the modulus in **Figure 4E**. Our thermal diodes are by far the softest among the ones reported in the literature (**Figure 1A**), which allows them to be folded, rolled, and conform well to rough surfaces, considerably expanding the range of applications. **Figure 4G** depicts the scenario in which the thermal diode is mounted on an uneven solid surface, where no voids or gaps at the interface can be observed. This is critical to ensuring good thermal contacts when the thermal diodes are used in real applications. The adhesion between the two terminals can also be important for the integrity of this material in applications. We found that the post-curing process can improve the adhesion strength. The adhesion strength measured by peeling tests shows the impact of post-curing time on adhesion which is the strength can be enhanced with a longer post curing time as shown in **Figure 4H**. During post-curation, the precursors that have not yet been fully cured in both terminals will continue to crosslink at the interface, strengthening the adhesion. For the thermal diode with an 80 wt.% alkane loading ratio, the adhesion strength can reach 0.06 N/cm after 60 min post-curing at 80 °C.

WAXD measurements are also performed on the C16/PU terminal with an 80 wt.% alkane loading ratio at temperatures above and below the melting point (**Figure 4I**). When the temperature is above the melting point, the result shows no peaks, indicating that the C16 is in the liquid phase without any crystallites. As the testing temperature drops below the melting point, several peaks can be detected at (010), (011), (200), and (111), suggesting that the crystal domains emerge in C16. Since heat can transfer more efficiently in the crystals,[46] it can lead to a larger effective thermal conductance in the forward direction.

We show this effect using MD simulations, where the steady state nonequilibrium method is applied to a heterojunction composed of C16 and C30 polymer blocks (the simulation details are provided in **SI**). **Figure 5A** displays a typical steady-state temperature profile.

The effective thermal conductivity ($k$) of the junction is calculated as $k = -\frac{Q \cdot L}{A \cdot \Delta T}$, $Q$ is the amount of rate of thermal energy transferred across the material, $L$ is the distance between the heat source and sink regions, $A$ is the cross-sectional area, and $\Delta T$ is the temperature difference between the heat source and sink. Both C16 and C30 blocks start in solid crystalline phases. With the forward temperature bias, the C16 portion is subject to temperature below its melting point, and thus its phase is still in solid crystal. Therefore, its thermal conductivity is high as indicated by the flat temperature profile. Since C30 has a higher melting point, its temperature is also flat despite being exposed to the high temperature side. Therefore, the effective thermal conductivity is high. Under the reverse temperature bias, the C16 portion is subject to temperature higher than its melting point, and thus it becomes amorphous liquid. Its thermal conductivity decreases as reflected by the high slope of the temperature profile. Since the temperature differences are the same across the junction in the forward and reverse cases, the heat flux is proportional to the effective thermal conductivity of the heterojunction. The ratio of the heat flux in these two cases thus equals to the thermal rectification factor, and this ratio is calculated to be 1.38 (**Figure 5B**). We note that in real samples, neither C16 nor wax are pure crystals as simulated in the MD simulations. However, the MD results show the possibility that thermal rectification is achievable via phase change induced by flipping the temperature bias.

As mentioned earlier, the thermal diode can be potentially used for building envelopes. We performed experiments on two plastic boxes (the thermal conductivity of boxes wall is about 0.2 W/mk): one is covered by envelop made of our thermal diode and another is bare one. These two boxes are placed in a heating environment with temperature profile shown in **Figure 6A**. The internal temperature of the bare box follows the environmental temperature change closely. However, for the box with thermal diode envelop, its internal temperature rise is significantly retarded compared to the environmental temperature rise, as the thermal diode serves as a thermal barrier when the environmental temperature is above the melting point of the C16 terminal. For the case of environmental temperature dropping down (**Figure 6B**), both boxes show similar cooling curves. This is because when the C16 terminal is exposed to lower temperatures, it freezes into solid and thus the thermal conductivity become high, allowing heat to transfer through more efficiently. These two experiments show the potential of our thermal diode in temperature regulation in buildings. Using the thermal conductivity of the thermal diode in forward and reverse cases, we model the indoor temperature changes by finite element analysis using real weather temperature database. **Figure 6C** illustrates the conditions the wall is subject to during nighttime and daytime. **Figure 6D** shows the predicted results of the indoor temperature for 96 continuous hours along with the environmental temperature. It is apparent that the use of the thermal diode dampens the temperature oscillation compared to the normal wall. The thermal diode helps hinder heat from entering the room during when the environmental temperature is high, but it behaves similarly to normal walls during nighttime when environmental temperature is low. Such a feature is quantified in **Figure 6E** where daytime and nighttime temperature differences between the environment and the room is presented.

## Conclusion

In summary, we have proposed and experimentally fabricated a thermal diode based on PCM-based composites made of alkanes and PU matrices. Guided by modeling results, we optimized the thermal diode design and achieved a rectification factor of 1.42. The thermal diode also features an ultra-soft mechanical behavior, with an elongation to failure over 300% and an elastic modulus of 0.4 KPa, allowing it to interface well thermally with rough solid surfaces. We have also experimentally shown its ability to retard the heating of an enclosure in a hot environment while does not impede enclosure cooling when environmental temperature drops. Our modeling also indicates that the thermal diode can be effective in preventing excessive heating of a room during the daytime and while allowing rapid cooling during nighttime, which makes it useful for buildings in hot climates.

## Experimental Section

*Materials*

Hydroxyl-terminated polybutadiene (HTPB, NISSO-PB GI-2000, number average molecular weight, Mn ~2000) was purchased from NIPPON SODA Co., Ltd. Hexamethylene diisocyanate based polyisocyanate (HDI-based polyisocyanate WANNATE® HT-100) was purchased from Wanhua Chemical Group Co., Ltd. 1,6-hexamethylene diisocyanate (HDI) and Hexadecane (C16, 99%) were obtained from Macklin Inc. Paraffin wax and Ditin butyl dilaurate (DBTDL, 95%) catalyst were obtained from Aladdin Inc. All materials were used as received without further processing.

*Thermal diode fabrication*

Thermal diode terminals were synthesized by one-pot polymerization from HDI, HDI-based polyisocyanate and HTPB with DBTDL catalyst. The detailed synthesis process is as follows: first, HDI (-NCO: 11.9 mmol/g, 0.84 g), HDI-based polyisocyanate (-NCO: 5.1 mmol/g 0.42 g) and HTPB (-OH: 1 mmol/g, 10 g) were mixed in a flask equipped with a magnetic stirrer. After stirring for 10 min at room temperature, weighted C16 or wax (50 wt.%, 60 wt.%, 70 wt.%, 80 wt.%) was added to the HDI/PHDI/HTPB mixture, respectively. The HDI/PHDI/HTPB/C16 and HDI/PHDI/HTPB/wax mixtures were obtained by raising the temperature to 80 °C and stirring for 10 min. Then, DBTDL (5 wt.‰) was added to the mixture while stirring for 2 min at 80 °C, and the mixture was poured into a polyethylene glycol terephthalate mold to polymerize at 80 °C for 4 h to obtain the wax/PU and C16/PU composites. Finally, wax/PU was stacked with C16/PU at 80 °C for 1 h, and the thermal diode with strong interfacial adhesion between the two terminals was obtained.

Through controlling the weight of the cured mixture in the mold, the thermal diode terminals with various thickness ratios were obtained. It is noted that, for the first set of thermal rectification measurement, the thickness ratio between the terminals was 1:1, while for the rest sets, the thickness ratio was chosen following the modeling results.

*Characterizations*

For the thermal rectification factor measurement, an in-house-built testing system was used, as shown in **Figure 2A**. This system consisted of a cold plate as the heat sink and a thermoelectric heater as the heat source. The thermal diode was placed between the heater and cooler to form a sandwich structure. The sides of the thermal diode were wrapped by thermal insulation materials to prevent heat leakage. A K-type thermocouple was placed between the thermal diode and heater to monitor the temperature variation at the heat source side. The cold plate as the heat sink is set to have a constant temperature. This system allows to measure the steady-state heat flux through the thermal diode at the preset temperature biases, as measured by thermocouple and the heater power. The measurement

details are as follows:

In the forward case, the C16/PU terminal was placed in contact with the cold plate. The cold plate was maintained at the temperature of $(T^* - \frac{1}{2}\Delta T)$ which was below the melting point, $T^*$, of the C16/PU terminal. We then adjust the power input to tune the temperature of the thermoelectric heater to the target value of $T^* + \frac{1}{2}\Delta T$. Then, the forward heat flux can be calculated using the power of the heater:

$$J_+ = \frac{U \cdot I}{A} \tag{4}$$

where $U$ and $I$ were the voltage and current respectively of the heater measured on the DC power supply, $A$ is the cross-sectional area of the thermal diode, which is 655.36 mm². For the reverse case, the thermal diode was simply flipped.

Mechanical property measurements were carried out using a Shimadzu Electronics Universal Testing Machine (AGX-V, Japan) at room temperature (25 °C). The samples were dumbbell-shaped specimens. The tension tests used a displacement rate of 3 mm/min. For the cyclic tension test, a stretching rate of 400%/min was used. Compress-stress experiments were conducted on a MAX-1kN-M-2 machine with a displacement rate of 1 mm/min. The hardness of square specimens (10.0 mm×10.0 mm) was measured using a shore hardness tester (HPEII-00, Bareiss, Germany). A digital microscope (VHX-7000N, Keynece) was used to image the interfacial contact between the thermal diode and the uneven aluminum surface. Interfacial adhesion force test was performed on an AGX-V instrument at a displacement rate of 1 mm/min. PU/C16 (30 mm × 5 mm × 1.5 mm) were adhered to PU/wax (30 mm × 5 mm × 1.5 mm) in 80 °C for different times (20 min, 40 min, 60 min), and the 90° peeling test was carried out as the inset of **Figure 4H** shows after the adhered sample with a displacement rate of 3 mm/min. The cryo-fractured surfaces of the composites were measured by an atomic force microscope (AFM, Dimension ICON, Bruker, USA) to characterize the matrix-inclusion interface, and the modulus at different locations was measured by the displacement curves as the probe scans from the PU matrix to the alkane. The surface energy ($\gamma_{PU}$) was measured on an OCA 20 Micro automatic contact angle measurement system, and the value was calculated by the Lifshitz-van der Waals/Lewis acid–base (LW–AB) approach.[47, 48] The contact angle test was recorded within 2 seconds after liquid dripping. The surface tension components ($\gamma_{PU}^{LW}$, $\gamma_{PU}^{+}$ and $\gamma_{PU}^{-}$) of the PU are calculated from measurement of the contact angles (**Figure S3**) using three different liquids (water, ethylene glycol and diiodomethane) whose surface tension components ($\gamma_{l}^{LW}$, $\gamma_{l}^{+}$ and $\gamma_{l}^{-}$) are known (**Table S2**).

$$\gamma_{PU} = \gamma_{PU}^{LW} + \gamma_{PU}^{AB} \qquad (5)$$

$$\gamma_{PU}^{AB} = 2\sqrt{\gamma_{PU}^{+} + \gamma_{PU}^{-}} \qquad (6)$$

$$\gamma_{PU}(cos\theta + 1) = 2\left(\sqrt{\gamma_{PU}^{LW}\gamma_{l}^{LW}} + \sqrt{\gamma_{PU}^{+}\gamma_{l}^{-}} + \sqrt{\gamma_{PU}^{-}\gamma_{l}^{+}}\right) \qquad (7)$$

where $\gamma^{LW}$ and $\gamma^{AB}$ are the Lifshitz-van der Waals (apolar) and Lewis acid–base (polar) surface energy components; $\gamma^{+}$ and $\gamma^{-}$ are the Lewis acid (electron-acceptor) and Lewis base (electron-donor) parameters of surface energy, respectively.

X-ray diffraction (XRD, D8 ADVANCE, Bruker, USA) were performed at 25 °C and 0 °C from 10° to 80° 2θ with a tube voltage of 40 kV and a tube current of 40 mA. Differential Scanning Calorimetry (DSC) measurements were performed on a DSC 2500 (TA

Instruments, USA) in $N_2$ with a temperature increment of 5 °C/min to determine the phase-transition temperature and calculate the enthalpy of the C16/PU composites with alkane loading ratios. The thermal conductivity of the C16, wax and PU were measured using a Hot Disk TPS 2500S measurement system. At least five samples were tested, and the results were averaged for each composition. The test data for materials is provided in **SI.**

For heating and cooling tests, two cubic plastic boxes are covered with thermal diode materials or not placed in an environmental chamber. A K-type thermocouple is placed at the center of the box to record the internal temperature as the chamber heats up or cools down from room temperature.

*MD Simulation*

All MD simulations were performed using the Large-scale Atomic/Molecular Massively Parallel Simulator (LAMMPS). The OPLS-AA potential, which has been shown to predict reasonable values of the thermomechanical properties of polyethylene,[49] was used to model the PE (C16 and C30). The interactions between PE chains were van der Waals interactions, which were modeled by Lenard Jones (L-J) potentials.[50, 51] The relevant parameters are listed in **Table S4**. An L-J force cutoff of 10 Å was used. The amorphous C16 ($C_{16}H_{34}$), crystalized C16 and C30 ($C_{30}H_{62}$) were built using the Material Studio software. The simulated models of the forward and reverse cases were obtained by constructing the junctions of crystallized C16 and C30, and amorphous C16 and crystallized C30, respectively. For each model, they were relaxed firstly in the NVT ensemble, and then run in the NVE ensemble to ensure the systems reach to thermal equilibrium. For the C16 with the preset crystalline phase, we adopted multi-step equilibration to keep the crystalline structure, while the chains in C16 amorphous phase were fully relaxed.[52] To calculate the temperature profile and heat flux in the forward and reverse cases, non-equilibrium molecular dynamic simulations (NEMD) were performed for the equilibrated systems.[53-55]

*Finite Element Analysis*

The finite element analysis was performed by the COMSOL Multiphysics software. The 1D finite element model was employed to simulate the transient conduction of a building wall with a thickness of 0.25 m considered to be a thermal diode, consisting of PCM and concrete layers (thickness ratio: 6/4) that exchange heat with outside and inside air through radiation and convection. By using the material thermal properties listed in **Table S5**, and the ambient temperature which can be obtained from the tabulated data provided by COMSOL, the internal temperature is predicted.

## Supporting Information

It includes the temperature profile calculated by the 1-D heat conductance model on the fabricated thermal diode with the 1:1 thickness ratio, the DSC measurement results, the contact angle measurement results, the thermal conductivity of the pure materials, the information about the theoretical model, the calculation method of surface energy, the parameters used in MD simulations, material thermal properties values used in finite element calculation, and supporting figures mentioned in the main text. This supporting information is available from corresponding authors.

## Acknowledgments

This work was supported by the National Natural Science Foundation of China (Grant No. 62104161 and 52073300), the Guangdong Province Key Field R&D Program Project (No.

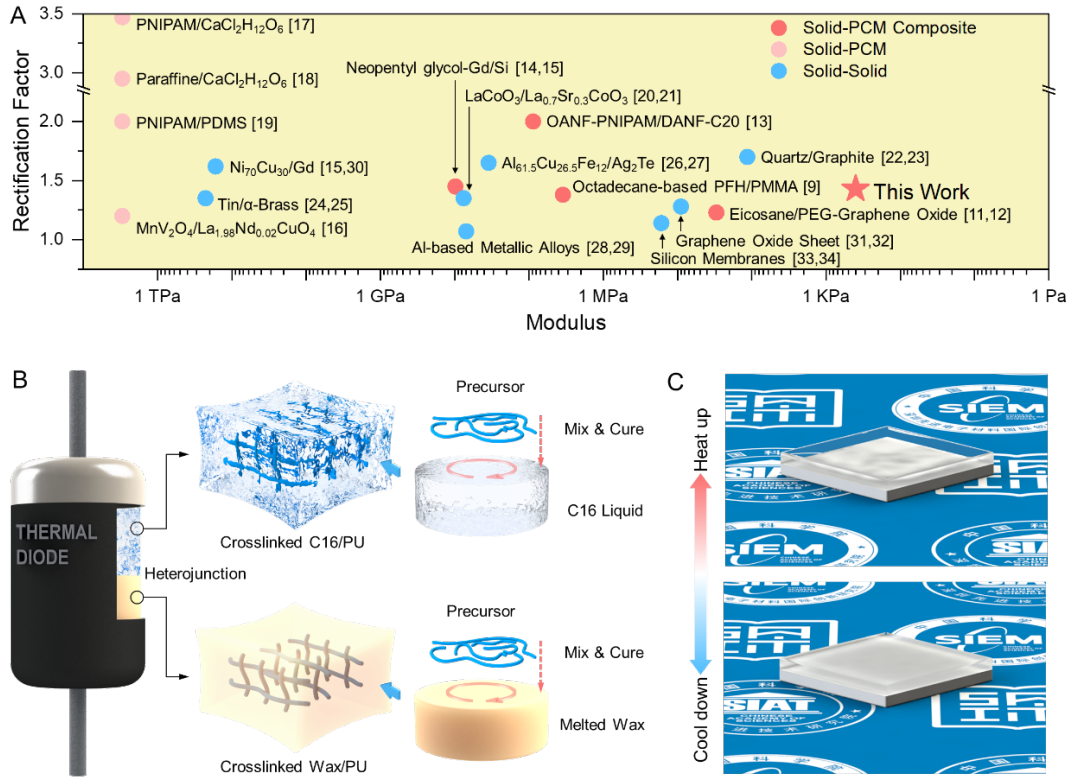

**Figure 1.** **(A)** Ashby chart of rectification factor and elastic modulus for the alkane-based phase change composites from this work compared to other heterojunction thermal diodes in the literature. The heterojunction types in the literature include solid-PCM composites,[9, 11-15] solid-PCM,[16-19] and solid-solid.[20-34] **(B)** The thermal diode of heterojunction is made of two terminals, C16/PU composite and wax/PU composite, which are obtained by crosslinking the PU precursor mixed with alkanes. **(C)** The phase of the C16/PU terminal will change when the environment temperature is heated up or cooled down across the melting point of C16 (18 °C).

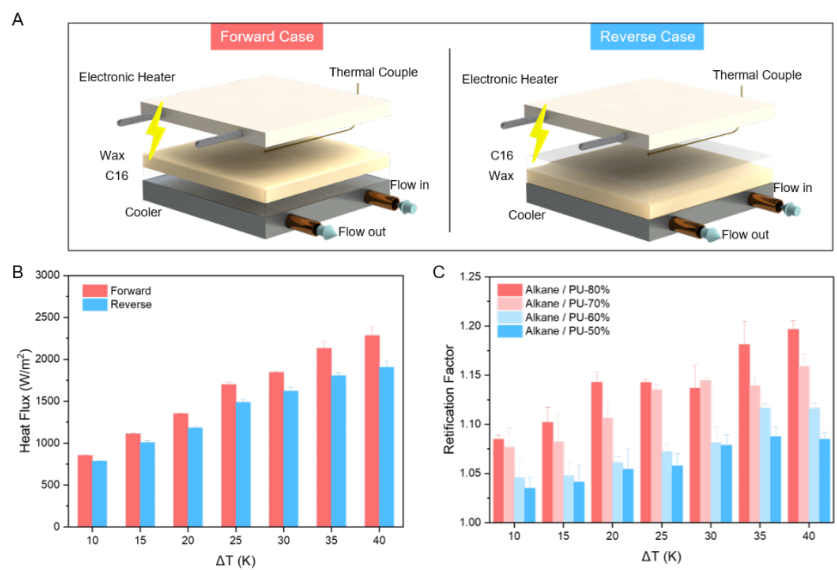

**Figure 2.** **(A)** The experimental setup for measuring the thermal rectification factor of the prepared thermal diode under the forward and reverse conditions. **(B)** Heat fluxes through the thermal diode with an 80% alkane loading fraction along the forward and reverse directions at different temperature biases. **(C)** The thermal rectification factor of the thermal diode with various alkane loading fractions at different temperature biases.

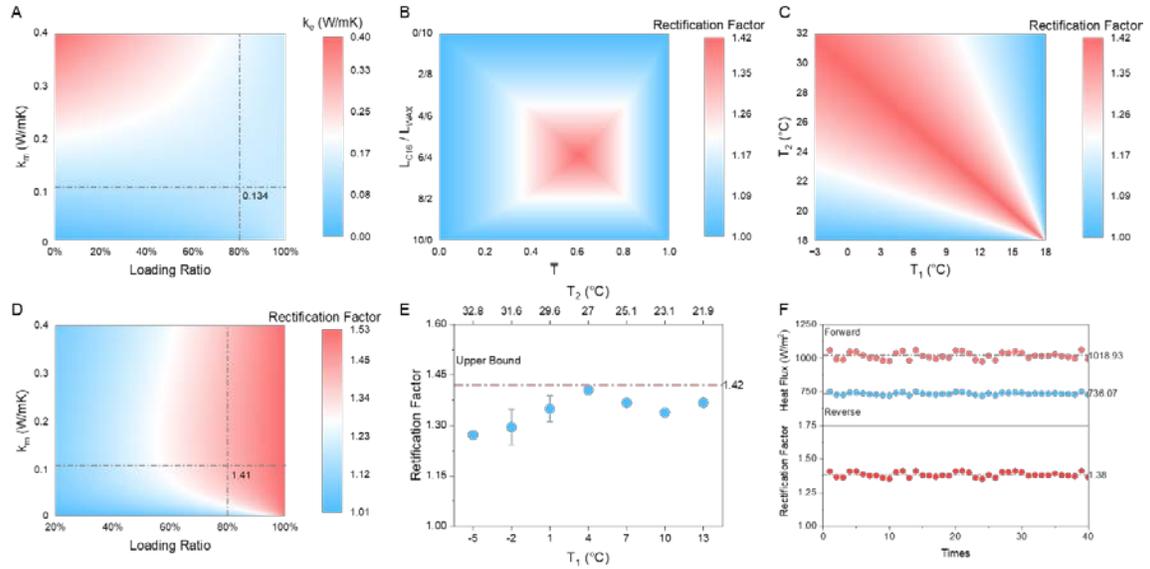

**Figure 3. (A)** The effective thermal conductivity ($k_e$) of alkane-based composite calculated by considering the loading ratio and the thermal conductivity ($k_m$) of the matrix material. The dashed line corresponds to $k_m = 0.11\ W/mK$ for the PU used as the matrix in our thermal diode. **(B)** The contour of calculated thermal rectification factor as a function of the reduced temperature and the terminal thickness ratio. **(C)** The thermal rectification factor as a function of the cold and hot side temperatures based on the optimal reduced temperature of 1.42 and terminal thickness ratio of 6/4 obtained in **(B)**. **(D)** The maximum predicted thermal rectification factor as a function of loading ratio and $k_m$. **(E)** The thermal rectification factors of the thermal diode with an 80% alkane loading ratio obtained with the optimized temperature pairs and terminals thickness ratio. **(F)** The cycling stability of the forward and reverse heat fluxes and thermal rectification factor under the temperature pair of 4 and 27 °C.

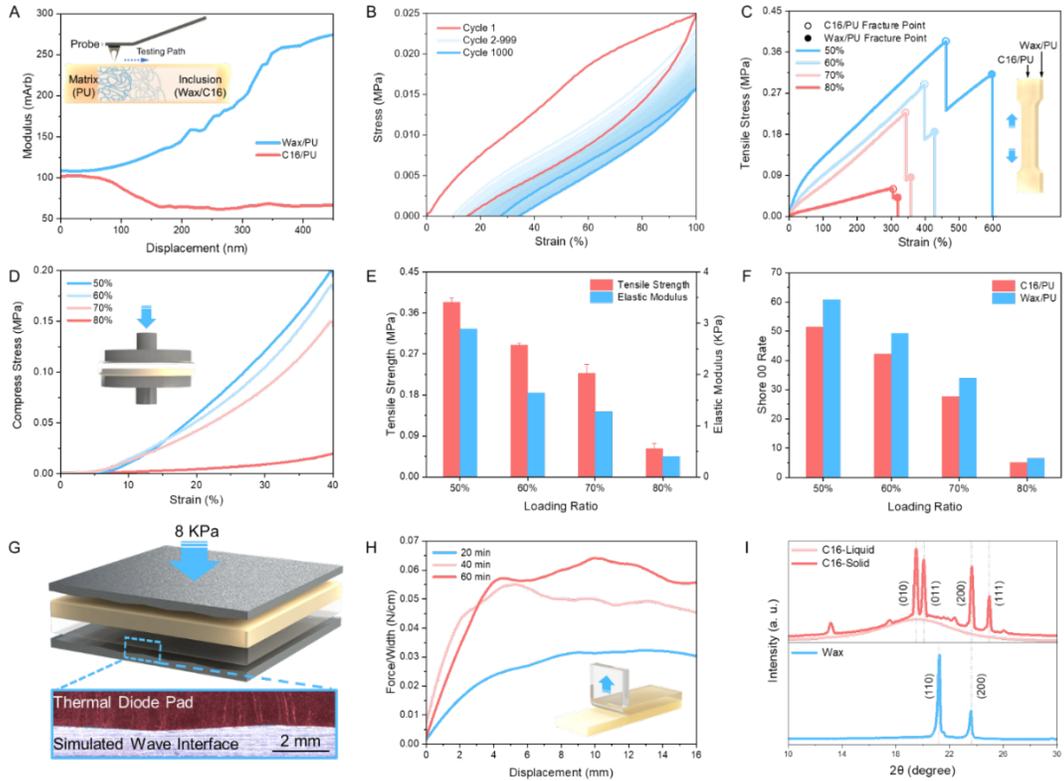

**Figure 4.** (**A**) AFM scan of stiffness along the illustrated path. The stiffness gradually transitions from that of PU to that of alkane. (**B**) 1000 cycles of tension test for the fabricated thermal diode with 80 wt.% alkane loading. (**C**) Strain-stress curves of thermal diodes with various alkane loading ratios under tension tests and (**D**) compression tests. (**E**) Tensile strength and elastic modulus of thermal diodes with various alkane loading fractions. (**F**) Stiffness of thermal diodes with various loading fractions. (**G**) The side view of the contact interface of the thermal diode and an uneven solid surface. (**H**) The interface adhesion between thermal diode terminals after post-curation with different curing times. (**I**) WAXD spectra of thermal diodes with 80 wt.% alkane loading ratio at different temperatures of 0 and 25 °C.

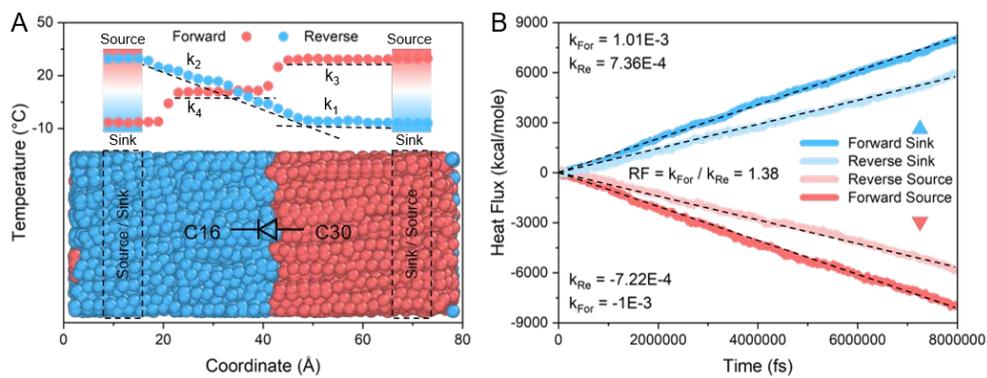

**Figure 5. (A)** The steady-state temperature profiles of the C16-C30 junction for the forward and reverse temperature biases. **(B)** Heat flux accumulation for both forward and reverse cases.

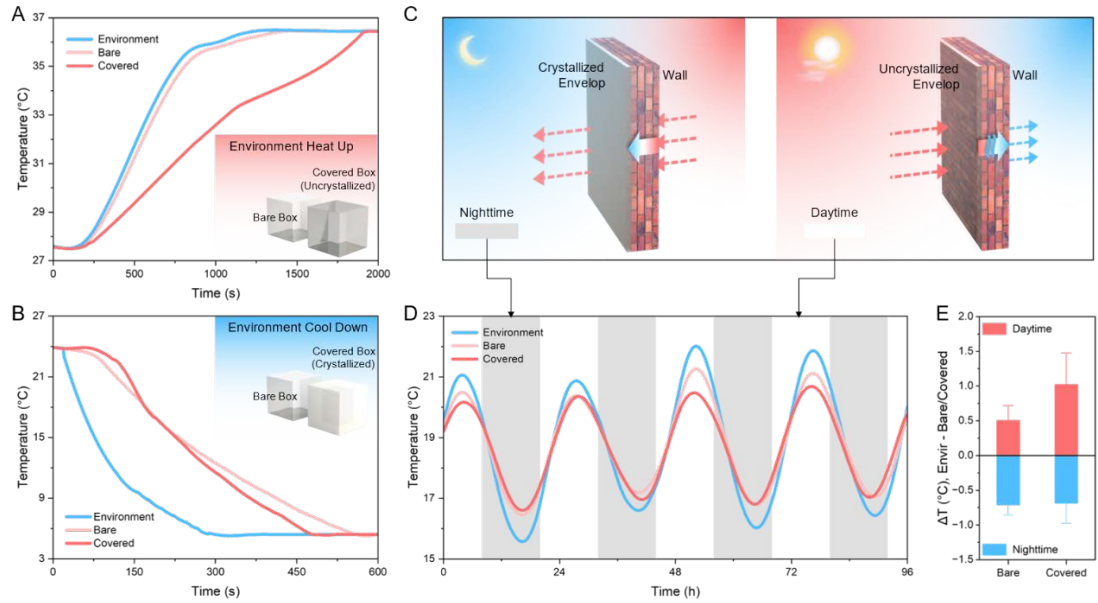

**Figure 6.** The temperatures of two boxes respectively enclosed by normal plastic and our thermal diode as walls when the environment (**A**) heats up and (**B**) cools down. (**C**) The schematic of the thermal diodes used as a building envelope during the daytime and nighttime. (**D**) The calculated temperature of a model room subject to real environmental temperature

fluctuations for 96 hours. (**E**) The temperature differences between the environment and the room during daytime and nighttime for rooms with and without the thermal diode covering the walls.

# Supporting Information

**Ultra-soft Thermal Diodes Enabled by Dual-Alkane-Based Phase Change Composites**

Yunsong Pang,[1]*# Junhong Li,[1]* Zhibin Wen,[1] Ting Liang,[2] Shan Gao,[1] Dezhao Huang,[3] Rong Sun,[1] Jianbin Xu,[2]# Tengfei Luo[4,5]# and Xiaoliang Zeng[1]#

[1]Shenzhen Institute of Advanced Electronic Materials, Shenzhen Institute of Advanced Technology, Chinese Academy of Sciences, Shenzhen 518055, China

[2]Department of Electronics Engineering, The Chinese University of Hong Kong, Shatin, N.T., Hong Kong 999077, China

[3]School of Power and Mechanical Engineering, Wuhan University, Wuhan, Hubei, 430072, China

[4]Department of Aerospace and Mechanical Engineering, University of Notre Dame, Notre Dame, Indiana 46556, United States

[5]Center for Sustainable Energy at Notre Dame, Notre Dame, Indiana 46556, United States

*These authors contributed to the work equally and should be regarded as co-first authors.

#Corresponding

Yunsong Pang, E-mail: ys.pang@siat.ac.cn,

Jianbin Xu, E-mail: jbxu@ee.cuhk.edu.hk

Xiaoliang Zeng, E-mail: xl.zeng@siat.ac.cn

Tengfei Luo, E-mail: tluo@nd.edu

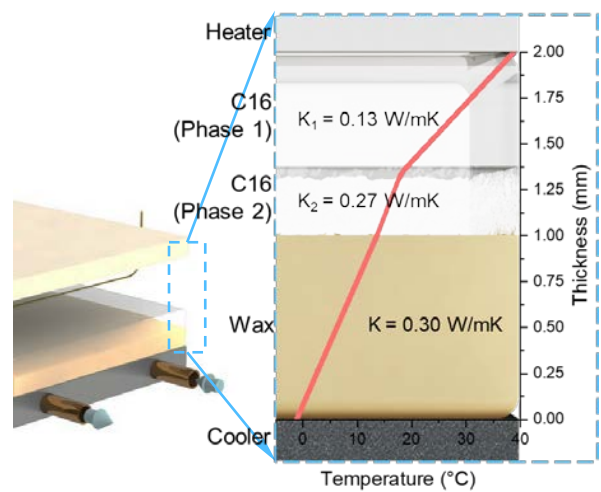

**Figure S1** The temperature profile for the thermal diode with a thickness ratio of 1:1, an upper surface temperature of 38 °C, and a lower surface temperature of -8 °C as estimated by a 1D heat transfer model.

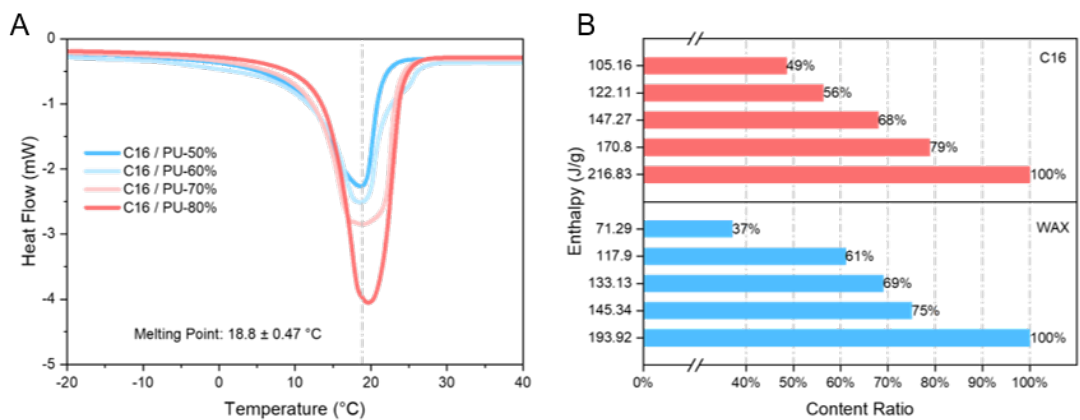

**Figure S2 (A)** The DSC results show that the phase transition points for the alkane/PU composite with various loading ratios are around 18.8 °C. **(B)** The enthalpy of composite materials is linearly correlated with the alkane loading fraction.

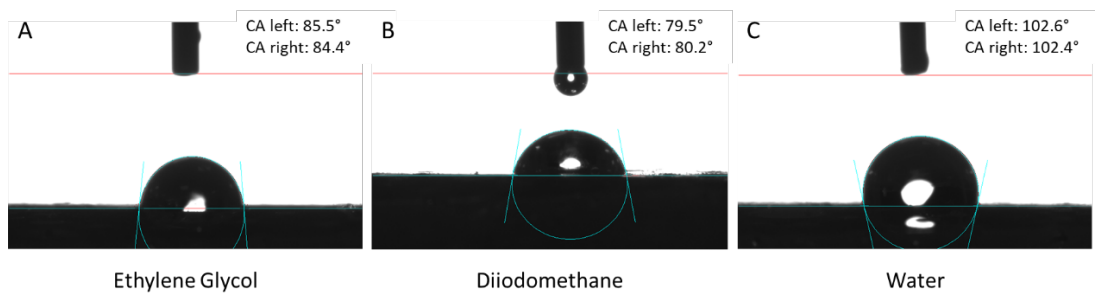

**Figure S3.** The contact angle of **(A)** ethylene glycol, **(B)** diiodomethane, and **(C)** water droplets on the surface of PU.

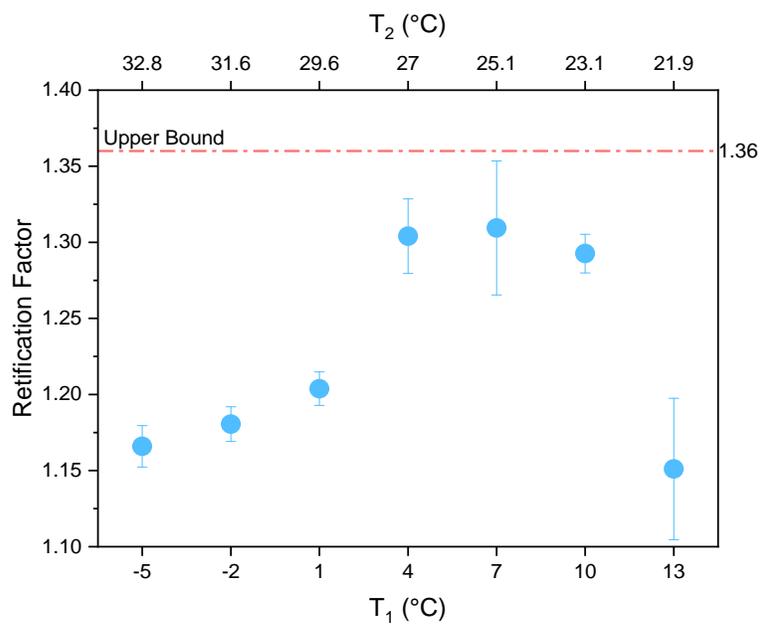

**Figure S4** Thermal rectification factors for thermal diodes with a 70% alkane loading fraction and 4:6 thickness ratio.

**Table S1.** The maximum thermal rectification factor ($\varepsilon$), the optimized terminals thickness ratio, and the suitable temperature pairs for the thermal diode can be estimated using the theoretical model from ref.[1]. In this table, $k_A$ refers to the constant thermal conductivity of material A, $k_{B,1}$ and $k_{B,2}$ refer to the phase change material B's thermal conductivity before and after phase transition, $L_A$ and $L_B$ are the thicknesses of the materials A and B, $T_1$ and $T_2$ are the temperatures of the heat sink and source, $T^*$ is material B's phase transition point.

|  | $\dfrac{k_A}{k_{B,1}} > \dfrac{k_A}{k_{B,2}}$ | $\dfrac{k_A}{k_{B,1}} \leq \dfrac{k_A}{k_{B,2}}$ |
|---|---|---|
| $\dfrac{L_A(T_2 - T^*)}{L_B(T^* - T_1)} > \dfrac{L_A(T_1 - T^*)}{L_B(T^* - T_2)}$ | $\varepsilon = \dfrac{k_A L_B + k_{B,1} L_A}{k_A L_B + k_{B,2} L_A}$ | $\varepsilon = \dfrac{k_{B,1}(T^* - T_1) + k_{B,2}(T_2 - T^*)}{k_{B,2}(T_2 - T_1)}$ |
| $\dfrac{L_A(T_2 - T^*)}{L_B(T^* - T_1)} \leq \dfrac{L_A(T_1 - T^*)}{L_B(T^* - T_2)}$ | $\varepsilon = \dfrac{k_{B,1}(T^* - T_1) + k_{B,2}(T_2 - T^*)}{k_{B,1}(T_2 - T_1)}$ | $\varepsilon = \dfrac{k_{B,2}(k_{B,1} L_A + k_A L_B)}{k_{B,1}(k_{B,2} L_A + k_A L_B)}$ |

**Table S2.** Surface energy components of the three probe liquids[2]

| Materials | $\gamma$ | $\gamma^{LW}$ | $\gamma^+$ | $\gamma^-$ |
|---|---|---|---|---|
| Water | 72.8 | 21.8 | 25.5 | 25.5 |
| Diiodomethane | 50.8 | 50.8 | ≈ 0.01 | ≈ 0 |
| Ethylene glycol | 48 | 29 | 3.0 | 30.1 |

**Table S3.** The thermal conductivity of each material measured by the hot disk method.

| Material | Thermal Conductivity, W/m·K |
|---|---|
| C16-liquid | 0.14 |
| C16-solid | 0.34 |
| Wax-solid | 0.38 |
| PU | 0.11 |

**Table S4.** Force field parameters for OPLS-AA. $k_\theta$, $k_n$, and $\epsilon$ are given in eV, $k_b$ in eV/Å$^2$, $r_0$ and $\sigma$ in Å.[3]

| Bond | $k_b$ | $r_0$ | $\epsilon$ | $\sigma$ |
|---|---|---|---|---|
| C-C | 23.2624 | 1.529 | 0.002864 | 3.5 |
| C-H | 29.512 | 1.09 | 0.001953 | 2.96 |
| H-h | - | - | 0.001302 | 2.5 |

| Angle | $k_\theta$ | $\theta_0$ |
|---|---|---|
| C-C-C | 5.1212 | 112.7° |
| C-C-H | 3.255 | 110.7° |
| H-C-H | 2.8644 | 107.8° |

| Dihedral | $k_1$ | $k_2$ | $k_3$ |
|---|---|---|---|
| C-C-C-C | 0.075516 | -0.00681 | 0.012109 |
| C-C-C-H | 0 | 0 | 0.015884 |
| H-C-C-H | 0 | 0 | 0.013801 |

**Table S5.** Thermal property values of material used in finite element analysis

|  | PCM layer | Concrete layer |
|---|---|---|
| **Density, $\rho$** | 770 kg/m$^3$ [4] | 1200 kg/m$^3$ * |
| **Heat Capacity, $C_p$** | 501.6 J/molK [4] | 1000 J/kgK * |
| **Thermal Conductivity, $k$** | 0.34 W/mK (T < 18.8 °C) #<br>0.14 W/mK (T > 18.8 °C) # | 0.2985 W/mK * |

# Provided from experimental results, * Provided from the Comsol database.